\pdfoutput=1
\documentclass{article}

\usepackage[T1]{fontenc}    
\usepackage{hyperref}       
\usepackage{url}            
   
\usepackage{amsfonts}
\usepackage{amsmath,amssymb}
\usepackage{mathtools}      

\usepackage{graphicx}
\usepackage{tabularx}
\usepackage{subcaption}

\usepackage[ruled, vlined, linesnumbered]{algorithm2e}

\title{Tie Strength in Online Social Networks and its Applications: A Brief Study}
   
\author{
  Chandni ~Saxena \\
  \texttt{cmooncs@gmail.com} 
   \and
Tanvir ~Ahmad\\
\texttt{tahmad2@jmi.ac.in } \\ 
\and 
 Department of Computer Engineering,
 \\Jamia Millia Islamia, 
 \\New Delhi-25, India }
\begin{document}
\maketitle 
\begin{abstract}
In online social network (OSN), understanding the factors bound to the role and strength of interaction(tie) are essential to model a wide variety of network-based applications. The recognition of these interactions can enhance the accuracy of link prediction, improve in ranking of dominants, reliability of recommendation and enhance targeted marketing as decision support system. In recent years, research interest on tie strength measures in OSN and its applications to diverse areas have increased, therefore it needs a comprehensive  review covering tie strength estimation systematically. The objective of this paper is to provide an in-depth review, analyze and explore the tie strength in online social networks. A methodical category for tie strength estimation techniques are discussed and analyzed in a wide variety of network types. Representative applications of tie strength estimation are also addressed. Finally, a set of future challenges of the tie strength in online social networks is discussed.
\end{abstract}




\section{Introduction}
\label{main}

\begin{figure}[!ht]
\centering
{\includegraphics[scale=0.32]{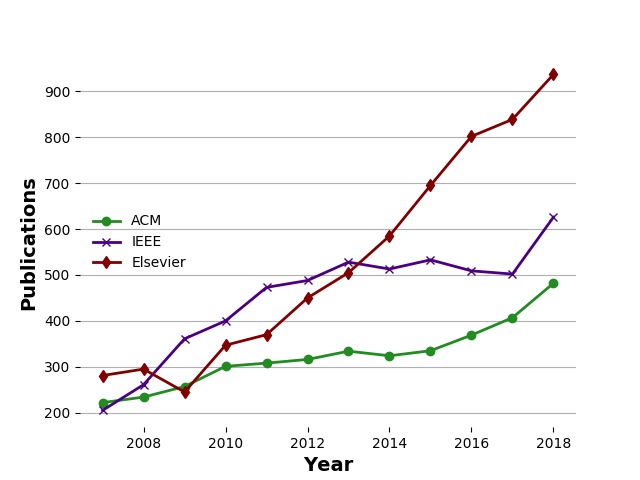}}
\caption{Publications growth for research on tie strength estimation in digital libraries.}
\label{fig0}
\end{figure}

A social network is a social organization of social actors and interactions among these actors as its integral parts. With the surge of information technology and online social networks (OSN) platforms such as Facebook, Instagram, Twitter, Linkedin, YouTube, Snapchat, Pinterest and 
ibo \cite{wiki2}; the collaboration between people have taken new dimensions and rise of online social networks. Meanwhile, splendid developments in mobile communication and wireless technology have improved the operations and services used by OSN, this resulted into emergence of another genre of social network called mobile social network (MSN). Essentially, MSN or OSN, have unique features where individuals with commonalities and alike interests can possibly connect people with one another. Social networks in these social context can be mapped into graphs, where nodes stand for individuals and the edges layout the relationships among them. The characteristics and features from the graph and other metrics related to such networks can be utilized to study the behavior of individual. Finally, such networks can leverage to define many interesting applications including  link prediction, recommendation system, ranking groups and individuals, pattern analysis and information diffusion \cite{zhou2009predicting,pappalardo2012well,larson2017weakness}.
The interactions and relationships among network structure can offer access to target information effectively. Knowledge of the strength of these relationships; furthermore the social dynamics provide new insights and shown to enhance correctedness of link prediction, reliability of recommendation  for collaboration, improve ranking for influence and fraud detection \cite{aghababaei2016mining,wang2015link}, make better model for information and disease spread and lead to enhance targeted marketing as decision support system. The property of relationship/tie strength ascertains to investigate and measure the interaction based on the knowledge extracted from the graph, social context, communication patterns and other related features. 
In the recent years, research interest on tie strength measure in online social networks and its applications to various areas has increased. Fig ~\ref{fig0} shows the importance of area field observed in terms of the number of published papers from three important digital libraries of computer science from 2007 to 2018 with search keywords ``tie strength online social networks''. The research trend on the problem of ties strength in online social networks has shown a rise during these years as increase in number of publications can be seen on this topic. In the light of this notable growth in area and to understand the wide variety factors related with tie strength, it is essential to cover the reviews study of the topic. This paper contributes an overview study of tie strength in online social networks and its applications to sort the solutions in various dimensions; including management science, business analytic, social science and computer science. First it gives the tie strength statement which includes formal definitions, types of ties and a generic framework for this metric. Then it presents strength estimation techniques from five major aspects such as; node based, communication pattern based, multimedia contents and learning based, geo-information based and temporal information based. It  majorly describes typical approaches based on topological features and multimedia contents based machine learning approaches with tie strength applications. Finally,  beside the current research, it outlines the future challenges for tie strength estimation. The rest of the paper is organized as follows. The problem statement and importance of tie strength is explained in section 2. Various techniques to estimate tie strength are classified and discussed in section 3. Section 4 covers range of applications of the  tie strength metric. Sections 5 and 6 draw the future challenges and conclusion from this study.   

\section{Problem Statement}
\begin{figure}
\centering
{\includegraphics[scale=0.32]{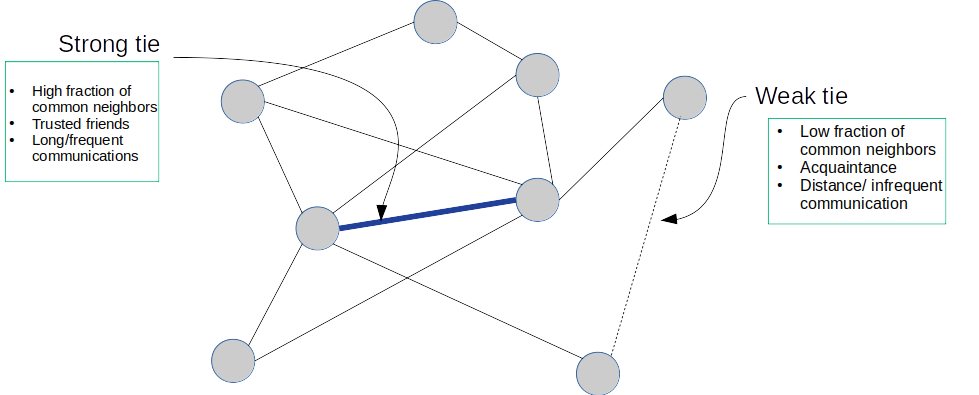}}
\caption{Example of strong and weak ties due to Granovetter's hypothesis.}
\label{fig1}
\end{figure}

Consider a social network $(G,V)$,  where $V$ denotes the set of nodes and $E$ denotes the set of links. The tie strength estimation aims to indicate the value ( nominal or discrete) of an edge between  two nodes which has a spectrum of widely accepted  definitions.  These ties in social networks have been categorized  in the seminal work by Granovetter \cite{granovetter1977strength} as weak ties and strong ties based on: interaction time, level of intimacy, emotional intensity and reciprocity. The strong ties are functional in a variety of situations with close friends and family members. Similarly, weak ties have less intense relationship between nodes according to above factors. Other class of ties such as latent ties, dormant ties and intermediate ties have also been defined by authors. The tie strength notion according to Granovetter's hypothesis (fig.~\ref{fig1}) can be explained from schematic network of nodes with related features to define weak and strong ties.
\subsection{Importance of strong ties and weak ties}The user in OSN can have weak or strong ties within the network, which maintains different levels of shared interest and belief system according to its strength.  In the detailed work of its kind, Onnela et al. \cite{onnela2007structure} studied the relationship between tie strength and structure of mobile networks on phone logs data. Authors found that strong ties cover prime role in maintaining local subgroups, however weak ties appear to cover important part for maintaining the networks cohesive strength. Weak ties are also related to the Burt's concept of structural hole \cite{burt2009structural} and  Granovetter's concept of local bridge \cite{granovetter1977strength} which play vital roles in  providing novel information  and information propagation on network. Strong ties are described to have strong control methods and tend to have trusted relationship which proved to target for network related decision support systems \cite{quijano2014development}.

\section{Tie strength estimation techniques} 
\begin{figure}
\centering
{\includegraphics[scale=0.30]{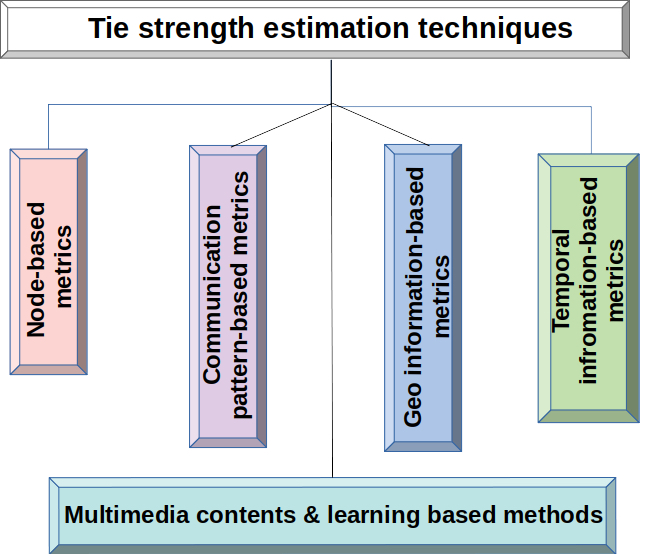}}
\caption{Category of tie strength estimation techniques on OSNs.}
\label{fig3}
\end{figure}

There are numerous fundamental approaches to estimate tie strength.The approaches mainly employ network topology, community, node and link level information. Approaches which look into the information related to social media data such as location and geographical information, co-location statistics, mobility behavior, communication patterns and temporal information, rely on the these platform and peer-to-peer communication patterns. Moreover, latent variables and learning based methods are more complex as authors incorporate additional external information for the estimation of this metric. This section tracks together a systematic review for the techniques of estimation of tie strength (fig.~\ref{fig3}) in online social networks.

\subsection{Node-based metrics}\label{AA}
Computing the tie strength of a node pair based on nodes information is an intuitive solution. According to ``\textit{the weak tie hypothesis}'' \cite{granovetter1977strength}, local network structure around pair of nodes exhibits important correlation with tie strength and relative overlap \cite{onnela2007structure} of common neighbors for nodes is defined as:
\begin{equation}
\centering
 O_{xy} = \frac{n_{xy}}{(k_x-1)+(k_y+1)-n_{xy}}
\label{eq1}
\end{equation}

Where $x$, $y$ connected nodes pair have $n_{xy}$ common neighbors and $k_x$, $k_y$ are individual neighbors of nodes $x$, $y$ respectively. Highly established node similarity indices such as path distance, number of common neighbors that two nodes share and  proximity of nodes pair \cite{liben2007link,zhou2009predicting} are used to infer link weight and tie strength measures which have further application to the link prediction problem. Newman~\cite{newman2001scientific} suggested a measure of tie strength of collaboration network which considers number of papers of two scientists in collaboration and number of coauthors for those collaborations as index of tie strength between them. Nowell and Kleinberg \cite{liben2007link} has reported  a review of tie strength measures based on local proximity of node pairs. Brand{\~a}o and Moro \cite{brandao2015analyzing} investigated co- authorship tie strength using common neighborhood overlap in the scholarly network. Node based metrics are basic and easy to incorporate as authors mainly use nodal information and actions, however additional attributes to this category leads to enhance  accuracy of estimating this metric \cite{kahanda2009using}.
\subsection{Communication pattern-based metric}
Online social networks provide IT-enabled communication environment, where social networks such as communities, society and organizations engage the individuals to communicate for the purposes and have certain interaction patterns. The communication in the forms of messages, emails and calls and all the contents of such interactions can be presented in terms of edge attributes. Therefore, analyzing the patterns of communication provides an interesting insight to estimate tie strength in such settings. Onnela et al.\cite{onnela2007structure} investigated tie strength metric by call duration and cumulative number of such calls between individuals along with its local topology in the mobile communication network. Pappalardo et al.~\cite{pappalardo2012well} explored a multidimensional network built over Facebook, Twitter and Foursquare to introduce tie strength estimation based on number on interactions of nodes.  Other studies \cite{wiese2014assessing,mattie2018understanding} have investigated accuracy of having communication patterns  correlated to the tie strength. Furthermore, incorporating the contents information of communication links can improve the precision of this metric.
\subsection{Geo information-based metrics}
Developement of web-based centralized applications and similarly functionality to mobile networks provide online social networks a deliverance of getting geographical. The evolution of geosocial networking \cite{wiki1} allows user to communicate relative to their locations. Amazon, Facebook, Twitter, Google and eBay are among such platforms to include social APIs and expand geolocation technology. The study of online social networks in this dimension has acknowledged property of these networks in relation to geographical attributes of social media users along with their topological structure \cite{valverde2018role}. The location data can be shared by users voluntarily or it can be based on geotagged contents such as tweets, flickr, Instagram or location based services like Foursquare check-ins. Geographical information provides additional features which is useful in applications such as location aware recommendation, marketing and sales analysis. Jeferrey et al.~\cite{mcgee2011geographic} investigated the positive correlation between tie strength and distance between two users on twitter. Pharm et al.~\cite{pham2013ebm} proposed an entropy-based model for estimating tie strength based on co-occurence of individuals. Sadilek et al.~\cite{sadilek2012finding} conferred location prediction approach based on social tie strength. Authors combined location-based features along with topology of friendship graph information to determine the social tie strength. Further related studies \cite{groh2013geographic,hsieh2015you} of measuring tie strength based on information such as distance between users, location and co-occurrence and their structural features, also confirm the positive dependency of  the metric on location-based user information. The geolocating applications and emerging volume of users location-based data demonstrate an obvious threat to the privacy issue in this regard. However, efficient and reliable use of such information can be utilized for cutting-edge business applications.
\subsection{Temporal information-based metrics}
Static interpretations of online social networks usually fail to embrace temporal aspects of human activity. Social relationships vary with time and tend to show human actions as Markovian and randomly distributed events against time \cite{miritello2013temporal}. The majority of OSNs show up this temporal pattern of changing edge sequences, varied intimacy of relations, emergence of new ties, time spans of communications, timelines of relationships and continuous changing activities over time. This temporal conduct of social networks can affect  other factors applicable to tie strength determination in online social networks. In result, structural property (clustering coefficient, neighborhood overlap) and  geographical property (community structure,distance, homophily) also receive conditional upshots \cite{miritello2013temporal}. With increasing availability of massive temporal information of how people act and communicate, facilitate to include temporal property of OSN to accurately characterize underlying characteristics of tie strength estimation and dynamic process of such networks. Karsai et al.~\cite{karsai2014time} conferred amount and time of interaction to define link  strength in large scale mobile phone network. Laurent et al.~\cite{laurent2015calls} defined strong tie strength based on network structure of mobile communication and frequency of interactions among intra-community nodes. Br{\~a}ndo et al.~\cite{brandao2017tie} identified tie strength in temporal network of co-authorship relations by measuring edge persistence and evolution of ties over time. To encode temporal network is a non-trivial task because it requires to maintain temporal ordering of edges and compute time with varying network properties. However, beyond this restraint of maintaining temporal features of OSN, tie strength estimation with temporal attributes could help in prominent applications such as co-authorship analysis and collaboration prediction, crime prediction in radically influenced networks \cite{aghababaei2016mining}, stock market fluctuation \cite{chen2014exploiting}, also ranking research\cite{freire2011ranking}. 
\subsection{Multimedia contents and learning-based tie strength}
\begin{figure}[htbp]
\centerline{\includegraphics[scale=0.32]{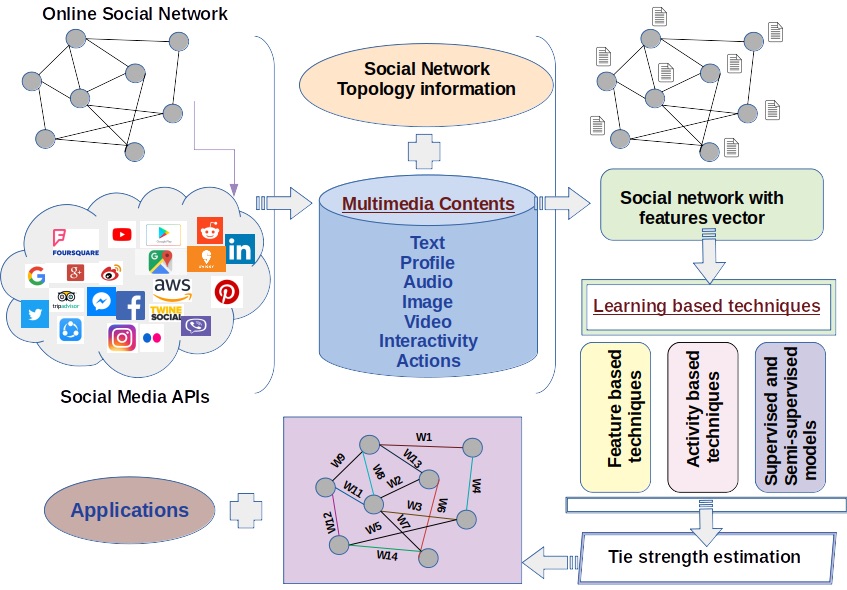}}
\caption{Tie strength estimation model on online social networks with multimedia contents.}
\label{fig2}
\end{figure}

A cascade growth in online plaforms of information system and social networking sites has shown a consequential surge in online content generation and sharing on such social networks. The contents generated by users include online comments on social networking sites such as Facebook and Twitter, sharing digital images (eg. on flickr), creating and working blogs, participating and posting on websites, creating tags, sharing ideas and sharing interests \cite{wiki1}. These contents are also called as linkage features or interaction contents of individuals, which provide schematic dimensions to the nodal attributes for a social network. The interaction contents of individuals on online social networks have been identified as strong factors for analyzing tie strength estimation \cite{zeng2013social,wang2013improved}.  A model view of real time update for online social networks and their tie strength estimation is represented in fig.~\ref{fig2}. The multimedia contents of user profiles are provided by social media through their APIs, where users participate for various purposes such as media posting, messaging, business, health, hobbies, shopping and geotagging. The attributed graphs with node's features vector are obtained from such user related contents. Further, applying various machine learning methods can determine the estimation of relationship strength on these attributed graphs.  The  learning methods for categorical node contents are provided further in this section.  
\subsubsection{Multivariate features-based learning methods}
Several recorded properties of online social networks play vital roles to understand various factors associated with tie strength estimation. Researchers have employed various methods of statistical machine learning for feature selection and prediction of key features for tie strength estimation \cite{mattie2018understanding,chen2011examination,wang2013improved,gilbert2009predicting} . Kahanda et al.~\cite{kahanda2009using} exploited 50 transnational features of friendship network from multidimensional graph for relationship estimation based on Naive Bayesian classifier, logistic regression and bagged decision tree models. Giebert and Karahalios ~\cite{gilbert2009predicting}  presented multivariate static model, based on classes of  user features to estimate tie strength on Facebook. Mattie et al.~\cite{mattie2018understanding} used random forest regression and classification to predict tie strength based on categorical features from CDR call network. He et al.~\cite{he2012principle} employed multivariate step-wise regression to determine key features for tie strength estimation and verified model on Naive Bayesian classifier. 
\subsubsection{User activity based learning methods}
The relationship strength among users in online networks varies in different activity fields therefore incorporating interaction activities with profile information can enumerate another dimension to relationship estimation. Zhao et al.~\cite{zhao2012relationship} proposed a framework to measure tie strength on Facebook network incorporating interaction among users in activity fields on OSN. Xiong et al.~\cite{xiong2016estimation} presented a model to measure tie strength between users in Sina network by considering similarity among profiles and interaction activity in various fields with co occurrence of users. Authors used graphical model for the approach. Abufouda and Zweig~\cite{abufouda2015we} addressed link estimation on  Facebook network, based on machine learning model on a set of associated interacting networks with Facebook's friends network.
\subsubsection{Supervised and semi-supervised learning models}
There is a wide spectrum of learning based link strength estimation models due to high variability in node features and data availability. Tang et al.~\cite{tang2011learning} proposed tie strength prediction task for partially labeled network data of mobile, email and publication networks. Authors proposed an algorithm to learn model parameters and to predict unknown relationships in a semi-supervised framework. Rotabi et al.~\cite{rotabi2017detecting} presented a supervised learning method of relationship strength estimation on twitter network. Authors used structural graph features from the presence and frequency of small graph motifs on combined weak and strong ties. Abufouda and Zweig~\cite{abufouda2017link} estimated social relationship strength based on friendship social network with external interaction using supervised learning model. Authors employed machine learning classification techniques to conduct the link evaluation via edge-proximity based label classification method. 

\subsection{Social theories-based metrics}
There exists numerous studies based on social theories for target of relationship strength estimation. A few reported studies are covered in this section. Sintos and Tsaparas~\cite{sintos2014using} exploited \textit{``Strong-Triadic-Closure''} Principle to characterize relationship strength on 4 different OSNs. Liberatore and Quijano-Sanchez~\cite{liberatore2017we} explored a questionnaire based validation and computational framework for tie strength determination using non-linear factor analysis models for various components related to social media. Pi et al.~\cite{pi2018inferring} considered mobile social network and proposed a unified framework to individual's inter relationship strength. Authors investigated time-aware mobility behavior, co-occurrence diversity, location semantic information, location significance and correlative impact on tie strength estimation. Adali et al.~\cite{adali2012actions} studied statistical features of communication patterns between users on social media. Authors considered reciprocity, attention, latency and assortativity as determinant features to capture contextual information as compared with textual features using Twitter data. Volkovich et al.~\cite{volkovich2012length} examined association between interaction strength, spatial inter-space and structural position of users relationships from \textit{Google's Tueti} communication network. Authors observed variation of tie strength with nodes k-core index, considering core and periphery structure of studied networks.
\section{Tie strength applications}
The tie strength metric can be used for diverse range of applications. This section remarks some typical applications including  link prediction, recommendation system, ranking groups and individuals, pattern analysis and information diffusion as follows.
\begin{itemize}         
\item \textit{Group recommendation:} The group recommendation is knowledge based system which provides recommendation to the group having social relationship to support decision making. The tie strength of group members are key factors to characterise and improve the quality of group recommendation system\cite{quijano2014development}.
\item \textit{Link prediction applications:} The social relationship strength estimation related to link prediction are among prominant techniques for such applications \cite{quijano2014development,zhou2009predicting,pappalardo2012well}.
\item \textit{Strategic game theoretic studies:} The weak ties play a prominant role in determining collective behaviour and maintaining cooperation in evolutionary process of prisoners dilema game \cite{xu2011importance}.
\item \textit{Location based friend prediction:} Increased availability to geographic social networks data and tie strength estimation techniques have made it straightforward to exploit location based social networks for its application to friendship prediction, user behaviour and location recommendation \cite{valverde2018role}.
\item \textit{Routing selection:} Social tie strength between nodes in mobile ad-hoc network can be exploited for improvement of optimized link state routing \cite{gupta2017routing}.
\item \textit{Collaboration prediction in scientific networks:} Tie strength estimation for coauthorship relationship network is a determinant factor for collaboration prediction in such networks \cite{newman2001scientific,brandao2015analyzing}.
\item \textit{Information diffusion and Network control:} The inclusion of weak ties in social networks can disrupt the spreading extent and speed of information, also exploited for network control in mobile social networks \cite{larson2017weakness,onnela2007structure}.
\item Other range of applications based on this metric includes crime prediction  \cite{aghababaei2016mining}, stock market fluctuation study \cite{chen2014exploiting}, friendship prediction \cite{valverde2018role,abufouda2015we} and pattern analysis \cite{pappalardo2012well}.
\end{itemize}   
\section{Future challenges}
Future work is imperative to improve and extend the link estimation precision. Online social networks are highly dynamic and a very few studies \cite{bapna2017trust} has been reported to deal with this issue. OSN tie strength prediction can be modeled using and incorporating dynamic online social networks.  A large number of methods for tie strength metric in OSNs consider only structural features and attributes, however social theory based methods have been investigated to limited number. More comprehensive work based on both approaches \cite{adali2012actions,volkovich2012length} can enhance the accuracy of the metric. The benchmark data sets for the study are inadequate to systematically appraise the performances and outline limitations of the models. The availability of benchmark data-sets for fair evaluation of model is next important step. In online social networks the incomplete or noisy data are usual conditions, therefore  more accurate tie estimation methods are required in such settings. Further, practical social networks with multiple relations types and heterogeneous nature of network need to challenge for the tie strength determination goals \cite{abufouda2015we,pi2018inferring}.
\section{Conclusion}
Tie strength estimation has been a primitive problem in the field of social organization and its applications are emerging with the advent of time due to advancement in technologies and availability of system data on online social networks.  This paper attempts to deliver a brief review of tie strength estimation techniques with variety of solutions from related researches. The fundamental concept of interaction strength is categorized and classified related to the state-of-the-art in tie strength computing. Tie strength metric estimation based  on topological features and users context related machine learning techniques are mainly presented in the paper.  Finally, the real time applications  are acknowledged and an excellent direction to future challenges are addressed which could resolve the current research issues related to tie strength computing.


\bibliography{m3}

\begin{thebibliography}{10}

\bibitem{abufouda2015we}
Mohammed Abufouda and Katharina~A Zweig.
\newblock Are we really friends?: Link assessment in social networks using
  multiple associated interaction networks.
\newblock In {\em Proceedings of the 24th International Conference on World
  Wide Web}, pages 771--776. ACM, 2015.

\bibitem{abufouda2017link}
Mohammed Abufouda and Katharina~A Zweig.
\newblock Link classification and tie strength ranking in online social
  networks with exogenous interaction networks.
\newblock {\em arXiv preprint arXiv:1708.04030}, 2017.

\bibitem{adali2012actions}
Sibel Adali, Fred Sisenda, and Malik Magdon-Ismail.
\newblock Actions speak as loud as words: Predicting relationships from social
  behavior data.
\newblock In {\em Proceedings of the 21st international conference on World
  Wide Web}, pages 689--698. ACM, 2012.

\bibitem{aghababaei2016mining}
Somayyeh Aghababaei and Masoud Makrehchi.
\newblock Mining social media content for crime prediction.
\newblock In {\em 2016 IEEE/WIC/ACM International Conference on Web
  Intelligence (WI)}, pages 526--531. IEEE, 2016.

\bibitem{bapna2017trust}
Ravi Bapna, Alok Gupta, Sarah Rice, and Arun Sundararajan.
\newblock Trust and the strength of ties in online social networks: An
  exploratory field experiment.
\newblock {\em MIS Quarterly}, 41(1):115--130, 2017.

\bibitem{brandao2017tie}
Michele~A Brand{\~a}o, Pedro~OS de~Melo, and Mirella~M Moro.
\newblock Tie strength dynamics over temporal co-authorship social networks.
\newblock In {\em Proceedings of the International Conference on Web
  Intelligence}, pages 306--313. ACM, 2017.

\bibitem{brandao2015analyzing}
Michele~A Brand{\~a}o and Mirella~M Moro.
\newblock Analyzing the strength of co-authorship ties with neighborhood
  overlap.
\newblock In {\em International Conference on Database and Expert Systems
  Applications}, pages 527--542. Springer, 2015.

\bibitem{burt2009structural}
Ronald~S Burt.
\newblock {\em Structural holes: The social structure of competition}.
\newblock Harvard university press, 2009.

\bibitem{chen2014exploiting}
Chen Chen, Wu~Dongxing, Hou Chunyan, and Yuan Xiaojie.
\newblock Exploiting social media for stock market prediction with
  factorization machine.
\newblock In {\em 2014 IEEE/WIC/ACM International Joint Conferences on Web
  Intelligence (WI) and Intelligent Agent Technologies (IAT)}, volume~2, pages
  142--149. IEEE, 2014.

\bibitem{chen2011examination}
Chien-Chou Chen, Terry Hui-Ye Chiu, Yuh-Jzer Joung, and Shy~Min Chen.
\newblock An examination of online social networks properties with
  tie-strength.
\newblock In {\em PACIS}, page~41, 2011.

\bibitem{freire2011ranking}
Vin{\'\i}cius~P Freire and Daniel~R Figueiredo.
\newblock Ranking in collaboration networks using a group based metric.
\newblock {\em Journal of the Brazilian Computer Society}, 17(4):255--266,
  2011.

\bibitem{gilbert2009predicting}
Eric Gilbert and Karrie Karahalios.
\newblock Predicting tie strength with social media.
\newblock In {\em Proceedings of the SIGCHI conference on human factors in
  computing systems}, pages 211--220. ACM, 2009.

\bibitem{granovetter1977strength}
Mark~S Granovetter.
\newblock The strength of weak ties.
\newblock In {\em Social networks}, pages 347--367. Elsevier, 1977.

\bibitem{groh2013geographic}
Georg Groh, Florian Straub, Johanna Eicher, and David Grob.
\newblock Geographic aspects of tie strength and value of information in social
  networking.
\newblock In {\em Proceedings of the 6th ACM SIGSPATIAL International Workshop
  on Location-Based Social Networks}, pages 1--10. ACM, 2013.

\bibitem{gupta2017routing}
Riten Gupta, Niyant Krishnamurthi, Uen-Tao Wang, Tejaswi Tamminedi, and Mario
  Gerla.
\newblock Routing in mobile ad-hoc networks using social tie strengths and
  mobility plans.
\newblock In {\em 2017 IEEE wireless communications and networking conference
  (WCNC)}, pages 1--6. IEEE, 2017.

\bibitem{he2012principle}
Yaxi He, Chunhong Zhang, and Yang Ji.
\newblock Principle features for tie strength estimation in micro-blog social
  network.
\newblock In {\em 2012 IEEE 12th International Conference on Computer and
  Information Technology}, pages 359--367. IEEE, 2012.

\bibitem{hsieh2015you}
Hsun-Ping Hsieh, Rui Yan, and Cheng-Te Li.
\newblock Where you go reveals who you know: Analyzing social ties from
  millions of footprints.
\newblock In {\em Proceedings of the 24th ACM International on Conference on
  Information and Knowledge Management}, pages 1839--1842. ACM, 2015.

\bibitem{kahanda2009using}
Indika Kahanda and Jennifer Neville.
\newblock Using transactional information to predict link strength in online
  social networks.
\newblock In {\em Third International AAAI Conference on Weblogs and Social
  Media}, 2009.

\bibitem{karsai2014time}
M{\'a}rton Karsai, Nicola Perra, and Alessandro Vespignani.
\newblock Time varying networks and the weakness of strong ties.
\newblock {\em Scientific reports}, 4:4001, 2014.

\bibitem{larson2017weakness}
Jennifer~M Larson.
\newblock The weakness of weak ties for novel information diffusion.
\newblock {\em Applied network science}, 2(1):14, 2017.

\bibitem{laurent2015calls}
Guillaume Laurent, Jari Saram{\"a}ki, and M{\'a}rton Karsai.
\newblock From calls to communities: a model for time-varying social networks.
\newblock {\em The European Physical Journal B}, 88(11):301, 2015.

\bibitem{liben2007link}
David Liben-Nowell and Jon Kleinberg.
\newblock The link-prediction problem for social networks.
\newblock {\em Journal of the American society for information science and
  technology}, 58(7):1019--1031, 2007.

\bibitem{liberatore2017we}
Federico Liberatore and L~Quijano-Sanchez.
\newblock What do we really need to compute the tie strength? an empirical
  study applied to social networks.
\newblock {\em Computer Communications}, 110:59--74, 2017.

\bibitem{mattie2018understanding}
Heather Mattie, Kenth Eng{\o}-Monsen, Rich Ling, and Jukka-Pekka Onnela.
\newblock Understanding tie strength in social networks using a local “bow
  tie” framework.
\newblock {\em Scientific reports}, 8(1):9349, 2018.

\bibitem{mcgee2011geographic}
Jeffrey McGee, James~A Caverlee, and Zhiyuan Cheng.
\newblock A geographic study of tie strength in social media.
\newblock In {\em Proceedings of the 20th ACM international conference on
  Information and knowledge management}, pages 2333--2336. ACM, 2011.

\bibitem{miritello2013temporal}
Giovanna Miritello.
\newblock {\em Temporal patterns of communication in social networks}.
\newblock Springer Science \& Business Media, 2013.

\bibitem{newman2001scientific}
Mark~EJ Newman.
\newblock Scientific collaboration networks. ii. shortest paths, weighted
  networks, and centrality.
\newblock {\em Physical review E}, 64(1):016132, 2001.

\bibitem{onnela2007structure}
J-P Onnela, Jari Saram{\"a}ki, Jorkki Hyv{\"o}nen, Gy{\"o}rgy Szab{\'o}, David
  Lazer, Kimmo Kaski, J{\'a}nos Kert{\'e}sz, and A-L Barab{\'a}si.
\newblock Structure and tie strengths in mobile communication networks.
\newblock {\em Proceedings of the national academy of sciences},
  104(18):7332--7336, 2007.

\bibitem{pappalardo2012well}
Luca Pappalardo, Giulio Rossetti, and Dino Pedreschi.
\newblock " how well do we know each other?" detecting tie strength in
  multidimensional social networks.
\newblock In {\em 2012 IEEE/ACM International Conference on Advances in Social
  Networks Analysis and Mining}, pages 1040--1045. IEEE, 2012.

\bibitem{pham2013ebm}
Huy Pham, Cyrus Shahabi, and Yan Liu.
\newblock Ebm: an entropy-based model to infer social strength from
  spatiotemporal data.
\newblock In {\em Proceedings of the 2013 ACM SIGMOD International Conference
  on Management of Data}, pages 265--276. ACM, 2013.

\bibitem{pi2018inferring}
Ting Pi, Lingwei Cao, Pin Lv, Zhili Ye, and Hao Wang.
\newblock Inferring implicit social ties in mobile social networks.
\newblock In {\em 2018 IEEE Wireless Communications and Networking Conference
  (WCNC)}, pages 1--6. IEEE, 2018.

\bibitem{quijano2014development}
Lara Quijano-S{\'a}nchez, Bel{\'e}n D{\'\i}az-Agudo, and Juan~A
  Recio-Garc{\'\i}a.
\newblock Development of a group recommender application in a social network.
\newblock {\em Knowledge-Based Systems}, 71:72--85, 2014.

\bibitem{rotabi2017detecting}
Rahmtin Rotabi, Krishna Kamath, Jon Kleinberg, and Aneesh Sharma.
\newblock Detecting strong ties using network motifs.
\newblock In {\em Proceedings of the 26th International Conference on World
  Wide Web Companion}, pages 983--992. International World Wide Web Conferences
  Steering Committee, 2017.

\bibitem{sadilek2012finding}
Adam Sadilek, Henry Kautz, and Jeffrey~P Bigham.
\newblock Finding your friends and following them to where you are.
\newblock In {\em Proceedings of the fifth ACM international conference on Web
  search and data mining}, pages 723--732. ACM, 2012.

\bibitem{sintos2014using}
Stavros Sintos and Panayiotis Tsaparas.
\newblock Using strong triadic closure to characterize ties in social networks.
\newblock In {\em Proceedings of the 20th ACM SIGKDD international conference
  on Knowledge discovery and data mining}, pages 1466--1475. ACM, 2014.

\bibitem{tang2011learning}
Wenbin Tang, Honglei Zhuang, and Jie Tang.
\newblock Learning to infer social ties in large networks.
\newblock In {\em Joint european conference on machine learning and knowledge
  discovery in databases}, pages 381--397. Springer, 2011.

\bibitem{valverde2018role}
Jorge~C Valverde-Rebaza, Mathieu Roche, Pascal Poncelet, and Alneu
  de~Andrade~Lopes.
\newblock The role of location and social strength for friendship prediction in
  location-based social networks.
\newblock {\em Information Processing \& Management}, 54(4):475--489, 2018.

\bibitem{volkovich2012length}
Yana Volkovich, Salvatore Scellato, David Laniado, Cecilia Mascolo, and Andreas
  Kaltenbrunner.
\newblock The length of bridge ties: structural and geographic properties of
  online social interactions.
\newblock In {\em Sixth International AAAI Conference on Weblogs and Social
  Media}, 2012.

\bibitem{wang2015link}
Peng Wang, BaoWen Xu, YuRong Wu, and XiaoYu Zhou.
\newblock Link prediction in social networks: the state-of-the-art.
\newblock {\em Science China Information Sciences}, 58(1):1--38, 2015.

\bibitem{wang2013improved}
Xinyu Wang, Chunhong Zhang, and Li~Sun.
\newblock An improved model for depression detection in micro-blog social
  network.
\newblock In {\em 2013 IEEE 13th International Conference on Data Mining
  Workshops}, pages 80--87. IEEE, 2013.

\bibitem{wiese2014assessing}
Jason Wiese, Jun-Ki Min, Jason~I Hong, and John Zimmerman.
\newblock Assessing call and sms logs as an indication of tie strength.
\newblock 2014.

\bibitem{wiki1}
{Wikipedia contributors}.
\newblock Geosocial networking --- {Wikipedia}{,} the free encyclopedia, 2018.
\newblock [Online; accessed 27-February-2019].

\bibitem{wiki2}
{Wikipedia contributors}.
\newblock Social media --- {Wikipedia}{,} the free encyclopedia.
\newblock
  \url{https://en.wikipedia.org/w/index.php?title=Social_media&oldid=885509949},
  2019.
\newblock [Online; accessed 1-March-2019].

\bibitem{xiong2016estimation}
Liyan Xiong, Yin Lei, Weichun Huang, Xiaohui Huang, and Maosheng Zhong.
\newblock An estimation model for social relationship strength based on users'
  profiles, co-occurrence and interaction activities.
\newblock {\em Neurocomputing}, 214:927--934, 2016.

\bibitem{xu2011importance}
Bo~Xu, Lu~Liu, and Weijia You.
\newblock Importance of tie strengths in the prisonerʼs dilemma game on social
  networks.
\newblock {\em Physics Letters A}, 375(24):2269--2273, 2011.

\bibitem{zeng2013social}
Xiaohua Zeng and Liyuan Wei.
\newblock Social ties and user content generation: Evidence from flickr.
\newblock {\em Information Systems Research}, 24(1):71--87, 2013.

\bibitem{zhao2012relationship}
Xiaojian Zhao, Jin Yuan, Guangda Li, Xiaoming Chen, and Zhoujun Li.
\newblock Relationship strength estimation for online social networks with the
  study on facebook.
\newblock {\em Neurocomputing}, 95:89--97, 2012.

\bibitem{zhou2009predicting}
Tao Zhou, Linyuan L{\"u}, and Yi-Cheng Zhang.
\newblock Predicting missing links via local information.
\newblock {\em The European Physical Journal B}, 71(4):623--630, 2009.

\end{thebibliography}
\bibliographystyle{plain}
\end{document}